\documentclass[letterpaper,twocolumn,prl,aps,superscriptaddress,amsmath,amssymb,floatfix]{revtex4-1}
\usepackage{mathptmx}
\usepackage[latin9]{inputenc}
\usepackage{color}
\usepackage{amsmath}
\usepackage{amssymb}
\usepackage{graphicx}
\usepackage{esint}
\usepackage[unicode=true,bookmarks=true,bookmarksnumbered=false,bookmarksopen=false,breaklinks=false,pdfborder={0 0 1},backref=false,colorlinks=true]{hyperref}
\hypersetup{linkcolor=magenta,urlcolor=blue,citecolor=blue,pdfstartview={FitH},hyperfootnotes=false}

\makeatletter

\usepackage{textcomp}
\usepackage{epstopdf}

\pdfpageheight\paperheight
\pdfpagewidth\paperwidth

\@ifundefined{textcolor}{}{%
 \definecolor{BLACK}{gray}{0}
 \definecolor{WHITE}{gray}{1}
 \definecolor{RED}{rgb}{1,0,0}
 \definecolor{GREEN}{rgb}{0,1,0}
 \definecolor{BLUE}{rgb}{0,0,1}
 \definecolor{CYAN}{cmyk}{1,0,0,0}
 \definecolor{MAGENTA}{cmyk}{0,1,0,0}
 \definecolor{YELLOW}{cmyk}{0,0,1,0}
}

\usepackage{xcolor}\usepackage{soul}
\definecolor{blue}{rgb}{0,0,1}
\definecolor{red}{rgb}{1,0,0}
\definecolor{green}{rgb}{0,1,0}

\usepackage{soul}

\makeatother

\begin{document}
\title{Giant-atom quantum acoustodynamics in hybrid superconducting-phononic integrated circuits}

\author{Lintao~Xiao}
\thanks{These authors contributed equally to this work.}
\affiliation{Center for Quantum Information, Institute for Interdisciplinary Information
Sciences, Tsinghua University, Beijing 100084, China}

\author{Bo~Zhang}
\thanks{These authors contributed equally to this work.}
\affiliation{Center for Quantum Information, Institute for Interdisciplinary Information
Sciences, Tsinghua University, Beijing 100084, China}

\author{Yu Zeng}
\affiliation{Laboratory of Quantum Information, University of Science and Technology of China, Hefei 230026, China}
\affiliation{Anhui Province Key Laboratory of Quantum Network, University of Science and Technology of China, Hefei 230026, China}

\author{Xiaoxuan~Pan}
\affiliation{Center for Quantum Information, Institute for Interdisciplinary Information
Sciences, Tsinghua University, Beijing 100084, China}

\author{Jia-Qi~Wang}
\affiliation{Laboratory of Quantum Information, University of Science and Technology of China, Hefei 230026, China}
\affiliation{Anhui Province Key Laboratory of Quantum Network, University of Science and Technology of China, Hefei 230026, China}

\author{Ziyue Hua}
\affiliation{Center for Quantum Information, Institute for Interdisciplinary Information Sciences, Tsinghua University, Beijing 100084, China}

\author{Hongwei Huang}
\affiliation{Center for Quantum Information, Institute for Interdisciplinary Information Sciences, Tsinghua University, Beijing 100084, China}

\author{Yifang Xu}
\affiliation{Center for Quantum Information, Institute for Interdisciplinary Information Sciences, Tsinghua University, Beijing 100084, China}

\author{Guangming~Xue}
\affiliation{Beijing Academy of Quantum Information Sciences, Beijing 100084, China}
\affiliation{Hefei National Laboratory, Hefei 230088, China}

\author{Haifeng~Yu}
\affiliation{Beijing Academy of Quantum Information Sciences, Beijing 100084, China}
\affiliation{Hefei National Laboratory, Hefei 230088, China}

\author{Xin-Biao~Xu}
\email{xbxuphys@ustc.edu.cn}
\affiliation{Laboratory of Quantum Information, University of Science and Technology of China, Hefei 230026, China}
\affiliation{Anhui Province Key Laboratory of Quantum Network, University of Science and Technology of China, Hefei 230026, China}

\author{Weiting~Wang}
\email{wangwt2020@mail.tsinghua.edu.cn}
\affiliation{Center for Quantum Information, Institute for Interdisciplinary Information
Sciences, Tsinghua University, Beijing 100084, China}

\author{Chang-Ling~Zou}
\email{clzou321@ustc.edu.cn}
\affiliation{Laboratory of Quantum Information, University of Science and Technology of China, Hefei 230026, China}
\affiliation{Anhui Province Key Laboratory of Quantum Network, University of Science and Technology of China, Hefei 230026, China}
\affiliation{CAS Center For Excellence in Quantum Information and Quantum Physics,
University of Science and Technology of China, Hefei, Anhui 230026, China}
\affiliation{Hefei National Laboratory, Hefei 230088, China}

\author{Luyan~Sun}
%\email{luyansun@tsinghua.edu.cn}
\affiliation{Center for Quantum Information, Institute for Interdisciplinary Information
Sciences, Tsinghua University, Beijing 100084, China}
\affiliation{Hefei National Laboratory, Hefei 230088, China}
%\date{\today}

\begin{abstract}
We demonstrate a giant atom by coupling a superconducting transmon qubit to a lithium niobate phononic waveguide at two points separated by about 600 acoustic wavelengths, with a propagation delay of 125\,ns. The giant atom yields non-Markovian relaxation dynamics characterized by phonon backflow and a frequency-dependent effective decay rate varying four-fold over merely 4\,MHz, corresponding to a Purcell factor exceeding 40. Exploiting this frequency-dependent dissipation, we prepare quantum superposition states with high purity. Our results establish phononic integrated circuits as a versatile platform for giant-atom physics, providing highly tunable quantum devices for advanced quantum information processing.
\end{abstract}
\maketitle

Over the past half century, the development of cavity quantum electrodynamics (QED) has profoundly advanced our understanding of fundamental light-matter interactions, providing an experimental platform for testing quantum optics phenomena such as the modification of spontaneous emission rates~\cite{Cavity-Enhanced,Reiserer2015,Haroche2020}. The cavity-enhanced strong coupling between single atoms and single photons establishes the basic building blocks for quantum information processing, enabling conditional phase gates between atoms and photons, photon-mediated atom-atom coupling, and atom-mediated photon-photon entanglement~\cite{Duan2004,Xubo2005PRA,Hassan2023PRA,Bastian2016Nature,Welte2018PRX,Grinkemeyer2025Science,Thomas2022PRX,Sharma2025PRR}. The extension of cavity QED to superconducting circuits and artificial atoms~\cite{Devoret2013,Clerk2020,Blais2021}, known as circuit QED, provides a scalable platform that has driven remarkable progress in quantum computing, including the first demonstrations of quantum computational advantage~\cite{Arute2019Nature,Wu2021} and the achievement of break-even quantum error correction~\cite{Ofek2016Nature, Sivak2023Nature,Ni2023Nature,Google2025Nature,Sun2025}.

Recently, quantum acoustics studying the coupling between phonon excitations in solids and two-level systems, has emerged as an active research frontier, establishing cavity quantum acoustodynamics (QAD)~\cite{Connell2010Nature,Martin2014Science} as an acoustic analog of cavity QED. The phononic cavities, employing the surface acoustic wave (SAW)~\cite{Martin2014Science,Manenti2017NatCommun,Satzinger2018Nature,Moores2018PRL,HQiao2023Science,Ruan2024} and bulk acoustic wave (BAW)~\cite{Yiwen2017Science,Yiwen2018Nature,HXu2022PRL,Bild2023Science,Yang2024Science} resonators, as well as suspended mechanical oscillator and phononic crystal devices~\cite{Connell2010Nature,Arriola2016PRA,Arriola2018PRX,Arriola2019Nature,MacCabe2020Science,Mirhosseini2020Nature,Cleland2022Nature,Bozkurt2023NatPhy,Bozkurt2025NatPhy}, have shown intriguing multi-mode strong coupling effects. Additionally, phononic integrated circuits have been proposed and demonstrated as a scalable platform for circuit QAD~\cite{Xu2025,wang_circuit_2025} by integrating superconducting qubits with phononic circuits. The Purcell effect, i.e., the phononic cavity-induced acceleration of spontaneous emission from superconducting qubits, is demonstrated, confirming that the fundamental light-matter interaction paradigm extends to the phononic circuits~\cite{wang_circuit_2025}.

Compared to circuit QED, phononic platforms offer distinctive advantages arising from the short acoustic wavelength and slow propagation speed, enabling exploration of giant-atom physics~\cite{GuoPRA2017,KockumPRL2018,Wang2021,Terradas2022PRA,Chen2025,Leonforte2025,Anonymous2025,Du2025}. A giant atom is characterized by a coupling region whose spatial extent is comparable to or larger than the wavelength of the flying carrier, a parameter regime that is extremely challenging to access in conventional cavity QED or circuit QED. Effective giant-atom systems can be realized by introducing multiple coupling points between the atom and the waveguide~\cite{Kockum2014PRA}, as demonstrated in circuit QED using centimeter-scale microwave resonators with simultaneous coupling at two spatially separated points~\cite{Kannan2020Nature}. Recent realizations using SAW have shown significant advantages~\cite{Andersson2019NatPhy}, achieving time delays between coupling points comparable to the qubit lifetime, which delay would require tens of meters of optical fiber or microwave cable in conventional cavity/circuit QED systems.

\begin{figure*}
\begin{centering}
\includegraphics[width=0.9\textwidth]{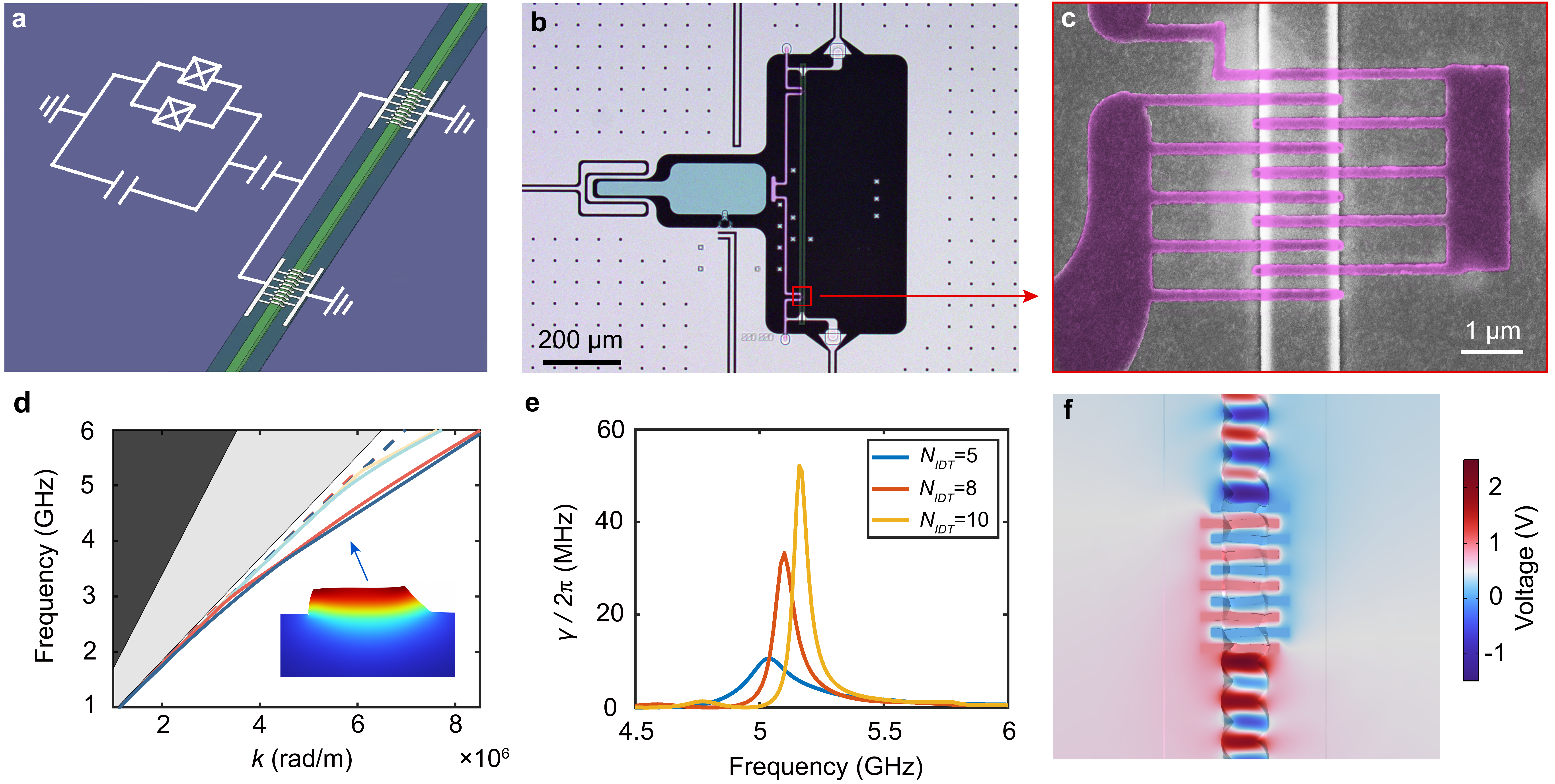}
\par\end{centering}
\caption{\textbf{Device for giant-atom quantum acoustodynamics.} \textbf{a}, Schematic of the hybrid superconducting-phononic system, consisting of a transmon qubit (white) coupled to a phononic waveguide (light green) via two interdigital transducers (IDTs). \textbf{b}, False-color optical photograph of the device. \textbf{c}, False-color scanning electron micrograph of an IDT. \textbf{d}, Dispersion curves of phononic modes in the phononic waveguide. Solid lines: guided modes; dashed lines: slab modes. Inset: displacement field of the fundamental quasi-Love mode. \textbf{e}, Simulated IDT coupling strength $\gamma$ versus qubit frequency for different numbers of finger pairs. \textbf{f}, Simulated electric field (red and blue colors) and displacement field (deformations) of the fundamental quasi-Love mode excited by an IDT with five pairs of fingers at 5~GHz.}
\label{Fig1}
\end{figure*}

In this Letter, we demonstrate a hybrid superconducting-phononic platform that couples a transmon qubit to a lithium niobate (LN) phononic waveguide~\cite{WeiFu2019NatCommun,Felix2021PRA,Xu2021APL,Wang2022APL,Balram2024APL} at two points separated by around 600 acoustic wavelengths, with a propagation delay of 125\,ns. This configuration realizes pronounced giant atom dynamics, including non-exponential decay and a frequency-dependent effective relaxation rate that varies four-fold over merely 4\,MHz, yielding a Purcell factor exceeding 40, substantially larger than previous SAW-based demonstrations. We further exploit this engineered dissipation landscape to implement bath engineering protocols~~\cite{Harrington2019PRA,Kitzman2023NatC} for deterministic preparation of quantum superposition states with high purity. Our platform bridges circuit QED and integrated phononic circuits~\cite{Xu2022c}, opening new avenues for quantum information processing with acoustic degrees of freedom.

Figures~\ref{Fig1}a-c illustrate the device, consisting of a frequency-tunable transmon qubit coupled to an LN waveguide fabricated on a sapphire substrate. Two interdigital transducers (IDTs), each comprising five pairs of metallic fingers, couple the qubit with the waveguide at two points separated by $L = 470 \,\mathrm{\mu m}$. The waveguide (width $w= 0.8\,\mathrm{\mu m}$ and thickness $t= 220\,\mathrm{nm}$) is fabricated with X-cut LN and aligned along the Y-axis to exploit the optimal piezoelectric coefficient~\cite{Felix2021PRA}. The contrast between the acoustic velocity in LN and the substrate leads to tight confinement of the phononic mode~\cite{WeiFu2019NatCommun,Felix2021PRA,Xu2021APL,Wang2022APL,Balram2024APL,Wang2025}. Figure~\ref{Fig1}d displays the dispersion curves of our waveguide, with the solid (dashed) lines indicating the Rayleigh and Love modes in the waveguide (slab). The fundamental quasi-Love mode [inset of Fig.~\ref{Fig1}d] is selected due to its high electromechanical coupling coefficient $k^2_{e}=6.3\%$, which far exceeds that of the quasi-Rayleigh mode ($k^2_{e}\approx 0.1\%$). At $5\,\mathrm{GHz}$, the corresponding IDT periodicity of $748\,\mathrm{n m}$ is fabricated for matching the wavelength of the quasi-Love mode. The IDT fingers are capacitively coupled to the qubit's floating pad, and the phononic continuum leads to an extra spontaneous emission rate of the qubits. As numerically calculated in Figs.~\ref{Fig1}e and \ref{Fig1}f, a coupling strength (defined as the extra decay rate) of $\gamma/2\pi = 10.8\,\mathrm{MHz}$ is predicted for $N_\mathrm{IDT}=5$ pairs of IDT fingers.

\begin{figure*}[t]
\begin{centering}
\textcolor{red}{\includegraphics[width=1\linewidth]{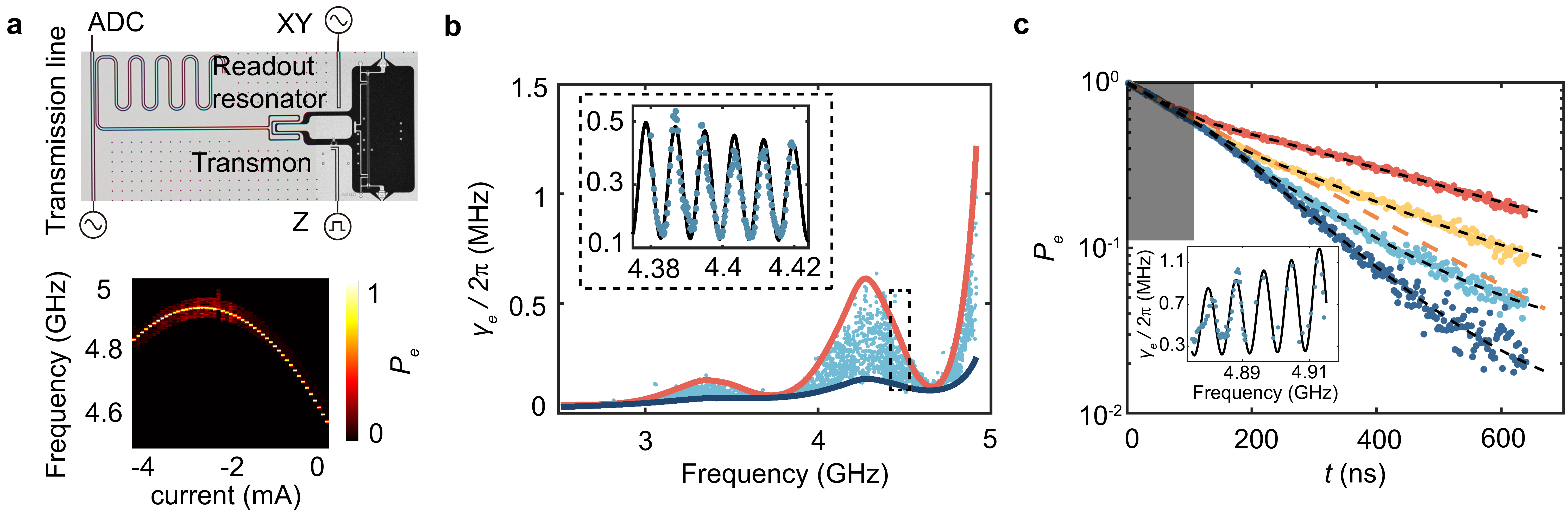}}
\par\end{centering}
\caption{\textbf{Giant atom relaxation dynamics.} \textbf{a}, Qubit readout. Upper panel: optical photograph of the circuit for qubit manipulation and readout. Lower panel: Spectrum of the measured qubit excitation probability $P_\mathrm{e}$ versus the bias current applied through the Z control line. \textbf{b}, Measured effective qubit relaxation rate $\gamma_e$ as a function of qubit frequency. Red and blue lines: fitted upper and lower envelopes. Inset: zoomed-in view of the modulation of $\gamma_e$; lines are fits. \textbf{c}, Qubit excitation decay dynamics at four qubit frequencies: $4.8887$ (blue), $4.8894$ (sky blue), $4.8899$ (yellow), and $4.8904\,\mathrm{GHz}$ (red). Solid curves: fits; yellow dashed line: extrapolation of the initial exponential decay for $0<t<T$ (shaded region) before interference occurs. Inset: expanded view of the modulation of $\gamma_e$ over the corresponding qubit frequency range.}
\label{Fig2}
\end{figure*}

The dual-point coupling geometry produces interference between phonons emitted from the two IDTs~\cite{Kockum2014PRA,GuoPRA2017,Andersson2019NatPhy}, leading to the periodic modulation of the qubit's decay rate as
\begin{equation}
\gamma_{e} = \gamma_\mathrm{in}+\gamma[1+\beta \mathrm{cos}(\omega T)],
\label{Eq1}
\end{equation}
where $\beta$ is the amplitude transmittance and $T$ is the delay due to the propagation of phonons emitted from one IDT to the other IDT, and $\gamma_\mathrm{in}$ is the qubit intrinsic decay rate. Figure~\ref{Fig2}a presents the setup to probe such a modulation of $\gamma_e$, where excitation pulses are applied to the qubit and the resulting excitation probability ($P_\mathrm{e}$) is measured to resolve its dynamics. Note that to suppress frequency-dependent excitation dynamics as studied later, we measure the effective relaxation rate of the qubit on an excitation timescale $t \gg T$, the same method used for measuring $T_1$. When scanning the bias current of the qubit, the qubit transition frequency ($\omega_{\mathrm{q}}$) varies, but a periodic suppression of excitation is observed, which has also been reported in SAW-based experiments~\cite{Andersson2019NatPhy}.

Direct measurement of the qubit decay rate $\gamma_e$, as shown in Fig.~\ref{Fig2}b, confirms the giant-atom effect predicted by Eq.~(\ref{Eq1}). The interference manifests as periodic oscillation of $\gamma_{e}$ in the inset of Fig.~\ref{Fig2}b, showing a period of about $8\,\mathrm{MHz}$ corresponding to a propagation time of $T=125\,\mathrm{ns}$, consistent with $L/v_g$ for a group velocity of $v_g =3600\,\mathrm{m/s}$ from simulation. Notably, we can switch the phonon emission of the qubit with a mere $4\,\mathrm{MHz}$ qubit frequency tuning, as shown by a four-fold enhancement of $\gamma_e$. The decay rate is fitted by Eq.~(\ref{Eq1}), with $\gamma_\mathrm{in}/2\pi=0.07\,\mathrm{MHz}$ and the oscillation visibility $\beta=0.78$. We also fit the envelopes of modulated $\gamma_e$ over a large frequency tuning range by $\gamma_{e} = \gamma_\mathrm{in}+\gamma(1\pm\beta)$, as shown by the red and blue lines, using the simulated IDT coupling strength $\gamma$ [Fig.~\ref{Fig1}e]. Here, $\gamma$ approximately follows a $\mathrm{sinc}^2$ function with distortions attributed to the reflections from the IDT fingers~\cite{Morgan2007Book}. We substitute a linearly increasing, frequency-dependent $\gamma_\mathrm{in}$ instead of a fixed value, accounting for the increased leakage loss from the IDTs to the bulky phononic continuum at short wavelengths. Comparing the observed decay rates $\gamma_e/2\pi = 1.11\,\mathrm{MHz}$ at $4.912\,\mathrm{GHz}$ and {$\gamma_e/2\pi = 27.2$ kHz} at $1.526\,\mathrm{GHz}$, the achieved Purcell factor exceeds $40$, substantially surpassing the value of 4.4 reported in previous SAW-based giant atom demonstrations~\cite{Andersson2019NatPhy}. Although an even larger coupling strength (and thus a higher Purcell factor) is predicted when qubit frequency exceeds $5.0\,\mathrm{GHz}$ [Fig.~\ref{Fig1}e and Fig.~\ref{Fig2}b], the achievable tuning range of the qubit is limited, as shown in Fig.~\ref{Fig2}a.

Beyond the modulation of the effective decay rate, giant atoms feature non-exponential relaxation dynamics, as shown in Fig.~\ref{Fig2}c. For four qubit frequencies near the maximal coupling, with accumulated phases $\mathrm{mod}[\omega T,\,2\pi]=0.164\pi,\,0.344\pi,\,0.484\pi$, and $0.601\pi$, respectively, the dynamics follow ~\cite{GuoPRA2017,Andersson2019NatPhy}
\begin{equation}
P_\mathrm{e}(t) = \sum_{n=0}^{\infty}\Theta(t-nT)\frac{(-\gamma \beta(t-nT))^n}{n!}e^{-i(\omega_\mathrm{q}-i\gamma-i\gamma_\mathrm{in})(t-nT)},\nonumber
\label{e3}
\end{equation}
as shown by the solid lines, where $\Theta(t)$ is the Heaviside function. An intuitive interpretation of this expression is that the excitation of the qubit has contributions from the phonon backflow, i.e., phonons emitted from one IDT couples back to the qubit after a propagation time $T$. Accounting for multiple rounds of backflow, the amplitude of the $n$-th backflow is suppressed by a factor of $\mathrm{exp}(-n\gamma T)$. Since our device parameter satisfies $\gamma T\sim 0.5$, these contributions of backflow are not negligible. The evolution exhibits two distinct regions separated by the delay time $T$. For $t<T$ (shade region), phonons emitted from one IDT have not yet reached the other, the backflow is prevented and the qubit decays exponentially. For $t>T$, the interference between  phonon emission and backflow causes the excitation decay deviate from simple exponential decay, as illustrated by the dashed line showing an extrapolation of the exponential decay. The experimental relaxation shows either suppressed or accelerated decay for destructive and constructive interference, exhibiting pronounced non-Markovian behaviors of the giant atom~\cite{Andersson2019NatPhy,GuoPRA2017}.

\begin{figure}
\begin{centering}
\textcolor{red}{\includegraphics[width=1\linewidth]{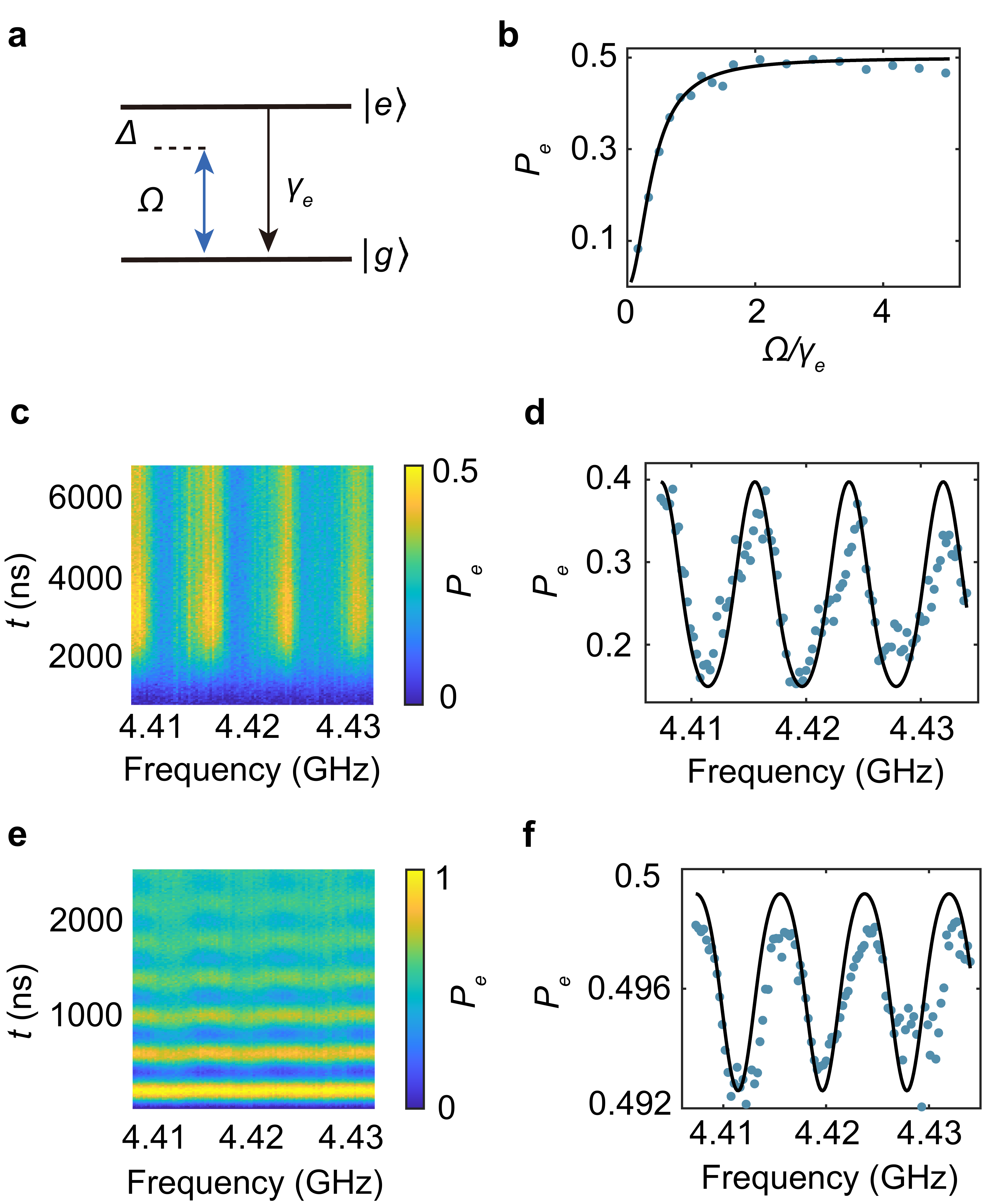}}
\par\end{centering}
\caption{\textbf{Driven giant atom dynamics.} \textbf{a}, Energy-level diagram of the driven qubit with Rabi frequency $\Omega$ and detuning $\Delta$. \textbf{b}, Steady-state qubit excitation probability $P_\mathrm{e}$ versus $\Omega$, with qubit frequency at $4.279\,\mathrm{GHz}$, $\Delta=0$, and a drive duration of $t=3.8\,\mathrm{\mu s}$ for long enough relaxation with $(\gamma_\mathrm{in}+\gamma)t\gg1$. \textbf{c}, Dynamics of $P_\mathrm{e}$ versus qubit frequency ($\Delta=0$) under a weak fixed $\Omega/2\pi =  0.2\,\mathrm{MHz}$. \textbf{d}, Corresponding steady-state $P_\mathrm{e}$. Solid line: prediction from the measured $\gamma_e$. \textbf{e} and \textbf{f}, Dynamic evolution (\textbf{e}) and steady-state $P_\mathrm{e}$ (\textbf{f}) under a strong drive with $\Omega/2\pi =2.5\,\mathrm{MHz}$.}
\label{Fig3}
\end{figure}

\begin{figure*}
\begin{centering}
\textcolor{red}{\includegraphics[width=1\linewidth]{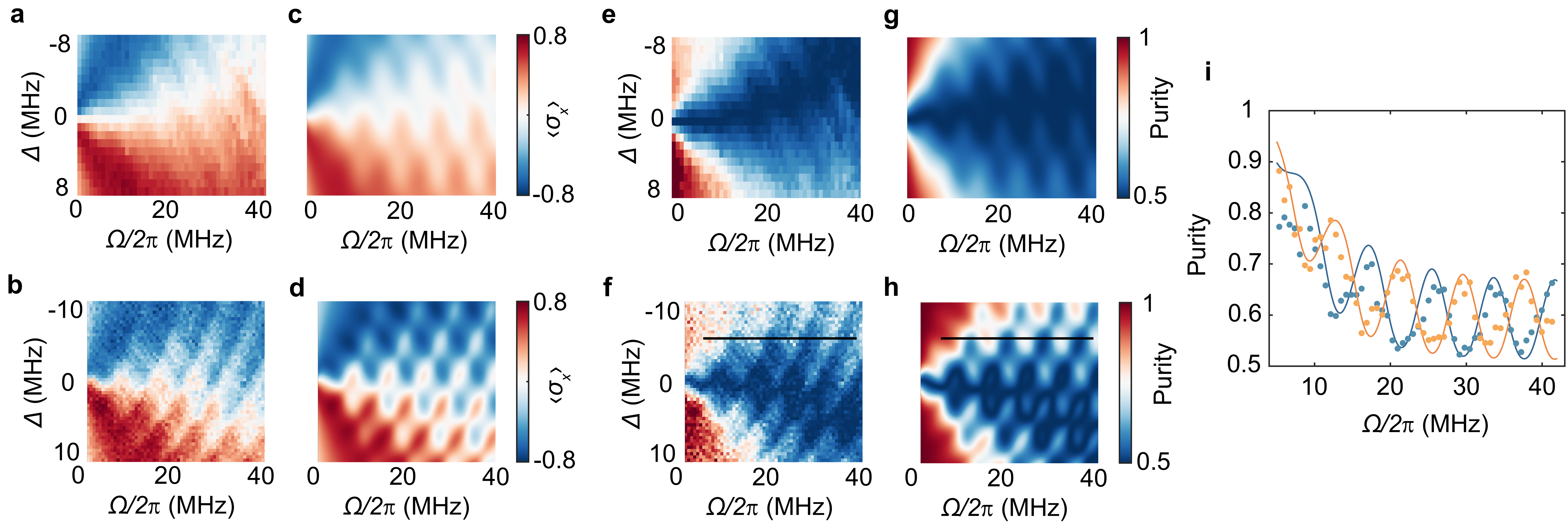}}
\par\end{centering}
\caption{\textbf{State preparation through relaxation of the driven giant atom.} \textbf{a} and \textbf{b}, Measured steady-State qubit coherence $\langle \sigma_x \rangle $ versus drive Rabi frequency $\Omega$ and detuning $\Delta$ at qubit frequencies of $4.640\,\mathrm{GHz}$ (\textbf{a}) and $4.891\,\mathrm{GHz}$ (\textbf{b}). \textbf{c} and \textbf{d}, Calculated $\langle \sigma_x \rangle $ for the same frequencies. \textbf{e} and \textbf{f}, Measured steady-state purity at qubit frequencies of $4.640\,\mathrm{GHz}$ (\textbf{e}) and $4.891\,\mathrm{GHz}$ (\textbf{f}). \textbf{g} and \textbf{h}, calculated steady-state purity. \textbf{i}, Steady-state purity versus $\Omega$ at $\Delta/2\pi=-5\,\mathrm{MHz}$ for two qubit frequencies: $4.891\,\mathrm{GHz}$ (blue) and $4.887\,\mathrm{GHz}$ (orange). Solid lines: theoretical results.}
\label{Fig4}
\end{figure*}

The giant atom not only exhibits non-Markovian relaxation dynamics, but also displays distinctive behavior under coherent drives due to the frequency-dependent coupling between the qubit and the continuum phonon bath in the waveguide. Figure~\ref{Fig3}a illustrates the energy levels of a giant atom driven by a microwave with a Rabi frequency $\Omega$ and a detuning $\Delta$ from the qubit transition frequency $\omega_\mathrm{q}$. First, we investigate the qubit excitation in the limit of long drive duration, where the decoherence effects erase the memory of the initial state and the population approaches a steady state.  As shown in Fig.~\ref{Fig3}b,  $P_\mathrm{e}$ is measured under varying $\Omega$ for $\Delta=0$ at $\omega_q/2\pi=4.279\,\mathrm{GHz}$. Neglecting the variation of the decay rate under driving, the steady-state solution of $P_\mathrm{e} = {\Omega^2}/({2\Omega^2+\gamma_{e}^2})$~\cite{Lindblad1976CMPhys,Foot2004Book} shows excellent agreement with the experimental data.

Under weak driving ($\Omega T\ll 2\pi$), the giant atom excitation dynamics for varying qubit frequencies are shown in Fig.~\ref{Fig3}c. With $\Omega/2\pi=0.2\,\mathrm{MHz}$, $P_\mathrm{e}$ increases with the drive duration before saturating at certain values, but also shows a periodic modulation with $\omega_\mathrm{q}$ as $P_\mathrm{e} \approx {\Omega^2}/{\gamma_{e}^2}$. As shown in Fig.~\ref{Fig3}d, the pronounced oscillations of the saturated $P_\mathrm{e}$ agree with the prediction from the giant atom's frequency-dependent decay rate extracted from Fig.~\ref{Fig2}b.

When the drive amplitude increases to $\Omega/2\pi=2.5\,\mathrm{MHz}$ ($\Omega T \approx 2\pi$), the driven giant atom enters a quantitatively different regime, with the excitation dynamics shown in Fig.~\ref{Fig3}e. In contrast to the modulation of $P_\mathrm{e}$ along the frequency axis, the giant atom in the strong-driving limit ($\Omega\gg\gamma_e$) shows modulation along the time axis due to Rabi oscillations. In this regime, the steady-state $P_\mathrm{e}\approx1/2$ regardless of the specific value of $\gamma_e$ and shows negligible frequency dependence. Only a residual oscillation with an amplitude around $0.008$ persists, which is just $4\%$ of the oscillation contrast observed in the weak-driving case.

In the strong-driving limit, new physical insight arises from the inherent nonlinearity of the two-level giant atom. For a detuned drive ($\Delta\neq0$), the qubit forms two dressed states with non-degenerate frequencies ($\pm\sqrt{\Omega^2+\Delta^2}$ with respect to the drive frequency $\omega$). The frequency-dependent dissipation then induces unbalanced decays of these dressed states, providing a knob to engineer the steady state. For a conventional qubit, symmetric dissipation rates of the dressed states eventually relax the driven qubit to a maximally mixed state. In contrast, the imbalanced dissipation rates in a giant atom will relax the system to one preferred dressed state, yielding a steady state with finite coherence~\cite{Harrington2019PRA,Kitzman2023NatC}.

Figure~\ref{Fig4} demonstrates this state preparation protocol at two qubit frequencies ($\omega_\mathrm{q}/2\pi=4.640\,\mathrm{GHz}$ and $4.891\,\mathrm{GHz}$) for varying drive detunings $\Delta$ and Rabi frequencies $\Omega$. Figures~\ref{Fig4}a and \ref{Fig4}b present the observed steady-state coherence, defined as the expectation value of the qubit Pauli matrix $\langle \sigma_x \rangle$. For a drive duration of $3.8\,\mathrm{\mu s}$, the numerical results are shown in Figs.~\ref{Fig4}c and \ref{Fig4}d according to the Lindblad master equations~\cite{Harrington2019PRA,Kitzman2023NatC,Lindblad1976CMPhys}. The agreement between experiment and theory is excellent. At $4.640\,\mathrm{GHz}$, although the IDT coupling strength $\gamma$ is near its minimum, the interference between the two coupling points still induces frequency-dependent modulation of $\gamma_e$ and generates coherence. The coherence of the steady state is evaluated by its purity, which is calculated by reconstructing the reduced density matrix through tomography. The experimental results are presented in Fig.~\ref{Fig4}e. As also confirmed by the numerical results [Fig.~\ref{Fig4}g], high purity is generated with large detunings and small drive amplitudes, as the qubit remains mostly in the ground state. However, even at very large drive amplitudes, we still observe high purity $\sim0.75$, indicating the effective preparation of a coherent steady state with the unbalanced dissipation of the giant atom's dressed states. At $4.891\,\mathrm{GHz}$, where the IDT coupling strength $\gamma$ is stronger and the modulation of $\gamma_e$ is more significant, the corresponding non-trivial coherence and purity in the driven giant atom are more pronounced, as shown in Figs.~\ref{Fig4}b and \ref{Fig4}f.

The detailed signature of purity is examined in Fig.~\ref{Fig4}i, where its dependence on $\Omega$ is studied for a fixed $\Delta/2\pi=-5\,\mathrm{MHz}$, corresponding to the red cut lines in Figs.~\ref{Fig4}f and \ref{Fig4}h. The purity exhibits periodic oscillations against $\Omega$, with a period of $8\,\mathrm{MHz}$. This period matches the inherent periodic modulation of the decay rate of the giant atom, which is anticipated from the periodic modulation of the dressed state sideband dissipation by scanning their frequencies $\sim\omega\pm\Omega$ ($\Omega\gg\Delta$) for increasing $\Omega$. Additionally, shifting the qubit frequency by $4\,\mathrm{MHz}$, i.e., half the decay modulation period of the giant atom, we can approximately flip the sign of the derivative $d\gamma_e/d\omega$, thereby inverting the decay rates of two dressed states. Consequently, we observe an anti-synchronized oscillation of purity, as shown by the orange dots and line in Fig.~\ref{Fig4}i. These results validate the unique property of the giant atom to prepare high-purity steady states across a broad parameter range by simply adjusting the qubit and drive frequencies.

In conclusion, we have demonstrated a giant atom realized by coupling a superconducting qubit to a phononic waveguide at two spatially separated points, featuring slow acoustic propagation velocity and a strong waveguide Purcell factor exceeding 40. The dual-point coupling geometry produces phonon backflow, where excitations emitted from one coupling point return to re-excite the qubit via the other coupling point after a delay time of 125 ns. This leads to pronounced frequency-dependent modulation of the effective relaxation rate and clear non-Markovian decay dynamics. Furthermore, we demonstrate that a strongly driven giant atom exhibits unbalanced decay of the dressed states, enabling deterministic preparation of high-purity steady states through the phononic continuum.

This hybrid superconducting-phononic platform provides a unique testbed for giant atom physics, accessing parameter regimes difficult or impossible to reach in conventional circuit QED or optical systems. The frequency-dependent dissipation landscape opens new routes for quantum state engineering and controlled single-phonon emission by merely tuning the qubit frequency, without requiring an auxiliary resonator. The platform can be further extended by enhancing the IDT coupling strength through an increased number of finger pairs and by scaling to multiple qubits or additional coupling points to realize more complex functionalities, including phonon-mediated entanglement and decoherence-free interactions~\cite{KockumPRL2018} for potential quantum information applications~\cite{Hann2019}.

\smallskip{}
\begin{acknowledgments}
This work was funded by Quantum Science and Technology-National Science and Technology Major Project (Grant Nos.~2024ZD0301500 and 2021ZD0300200) and the National Natural Science Foundation of China (Grant Nos.~92265210, 123B2068, 12104441, 12474498, 92165209, 12293053, 12374361, 92365301, and 92565301). The numerical calculations in this paper were performed on the supercomputing system in the Supercomputing Center of USTC, and this work was partially carried out at the USTC Center for Micro and Nanoscale Research and Fabrication.
\end{acknowledgments}

\end{document}